\documentclass[12pt]{article}
\usepackage{setspace,fullpage}
\newcommand {\tp}{t_p}

\begin{document}
\sloppy

\title{
On the Practicality of Using Intrinsic Reconfiguration As a Fault 
Recovery Method in Analog Systems
}

\author{
Garrison W. Greenwood \\
Department of Electrical \& Computer Engineering\\
Portland State University\\
Portland, OR 97207
}

\date{}

\maketitle

\bibliographystyle{unsrt}

\vspace{0.1in}


\vspace{0.1in}

\begin{abstract}
Evolvable hardware combines the powerful search capability of 
evolutionary algorithms with the flexibility of reprogrammable 
devices, thereby providing a natural framework for reconfiguration.  
This framework has generated an interest in using evolvable hardware 
for fault-tolerant systems because reconfiguration can effectively 
deal with hardware faults whenever it is impossible to provide 
spares.  But systems cannot tolerate faults indefinitely, which means 
reconfiguration does have a deadline.  The focus of previous evolvable 
hardware research relating to fault-tolerance has been primarily 
restricted to restoring functionality, with no real consideration of 
time constraints.  In this paper we are concerned with evolvable 
hardware performing reconfiguration under deadline constraints.  
In particular, we investigate reconfigurable hardware that undergoes 
intrinsic evolution.  We show that fault recovery done by intrinsic 
reconfiguration has some restrictions, which designers cannot ignore.
\end{abstract}

\section{Introduction}

Reliable systems can be depended on to provide continual service.  
Unfortunately, faults are inevitable, which leads to disruptions in service.  
One way of increasing a system's availability is to make it 
fault-tolerant---i.e., capable of detecting and recovering from hardware faults.  
Exchanging a faulty component with an operating spare is the most widely 
used method for hardware fault recovery \cite{aviz}, but it is not always 
possible to have redundant hardware available.  For instance, the very 
restrictive space and weight requirements typically found on spacecraft makes 
it difficult to find room for spares.  Reconfiguring a faulty system eliminates 
the need for redundant hardware, although reconfiguration does not always 
guarantee full functionality can be restored.  Nevertheless, reconfiguration 
is a viable fault recovery technique for any system with limited free 
space.  

Evolvable hardware (EHW) has emerged as a powerful method for doing 
original hardware design---which naturally suggests it could be equally useful for 
doing hardware reconfiguration.
The idea behind EHW is to combine the biologically-inspired 
search methods of evolutionary algorithms with the flexibility of 
reconfigurable hardware.  The evolutionary algorithm searches throughout the 
space of all possible configurations looking for the one that performs the best.  
Every configuration the evolutionary algorithm finds must be 
evaluated and there are two accepted methods: \textit{extrinsic evolution} 
where the evaluation is done in software, and \textit{intrinsic evolution} 
where the evaluation is done on a hardware implementation.  In many instances 
intrinsic evolution is necessary because the only real way to evaluate a 
configuration is to implement it and have it actually operate in the physical 
environment.  We refer to a reconfiguration search using intrinsic 
evolution as \textit{intrinsic reconfiguration}.   
(An excellent introduction to EHW can be found
in \cite{yao99}.) 

Two types of devices are suitable 
reconfigurable architectures for analog systems: 
the {field programmable analog array}
 (FPAA)
and the {field programmable transistor array}
 (FPTA).
These devices represent different levels of granularity.  The FPAA provides 
configurable blocks of circuitry along with programmable routing resources.  
Conversely, the FPTA consists of an array of MOSFET transistors interconnected 
via programmable switches.  A small number of capacitors are included on-chip, 
but resistors are synthesized using the MOSFET transistors.  FPAAs 
are available as commercial off-the-shelf (COTS) devices (e.g., see \cite{fpaa}); 
the FPTA was fabricated for NASA's Jet Propulsion 
Laboratory and is currently only available for research studies \cite{fpta1}.

Some recent work has shown EHW can be quite effective for reconfiguring 
existing hardware to overcome faults \cite{fpta2, mange, lukas}.   
The ability of evolutionary algorithms 
to find good reconfigurations is not at issue in this paper.  We instead
are concerned with  the issue of time.  Most EHW-based studies rely on device 
simulators rather than physical hardware.  This simulator 
use means the time to download a 
configuration, the time to program the device, and the time to 
conduct a fitness evaluation on the reconfigured  hardware has largely been 
ignored---even though they dramatically affect the evolutionary algorithm's 
running time.  

Systems cannot operate indefinitely with faults.  This means fault recovery 
must be completed within specified timeframes or undesirable consequences 
will occur.  The introduction of a deadline into fault recovery means 
fault-tolerant systems must be considered real-time systems (RTS).  
Greenwood, et.~al \cite{gree}
 were the first to suggest EHW-based reconfiguration 
has a time limit.  In this paper 
it will be shown how reconfiguration time 
impacts the choice of a fault recovery mechanism.

\section{Preliminary Definitions} \label{s2}

This section provides a brief introduction to real-time systems.  Interested 
readers are referred to some of the excellent books on this topic for further 
information (e.g., see \cite{burns}).  We begin with a formal definition of a 
real-time system.

\vspace{3ex}

Definition:  ({real-time system})

\vspace{2ex}

\textit{Any system that is both logically and temporally correct}

\vspace{2ex}

Logical correctness means the system satisfies all functional specifications.  
Temporal correctness means the system is guaranteed to perform these functions 
within explicit timeframes.  Fault-tolerant systems qualify as real-time systems 
because fault detection and fault recovery inherently have deadlines.  That is,
the fault must be detected within a certain period of time after it occurs, and
the fault must be corrected within a certain period of time after it is detected.  
Fault recovery may also have an expected start time.
	
The notion of real-time is often interpreted to mean really fast.  
This interpretation is not correct.  Real-time does not necessarily mean fast---and fast 
does not necessarily mean real-time.  Suppose a document must be sent from 
Chicago to London, and two delivery systems are available: surface mail 
with a guaranteed 3-day delivery time or e-mail with a guaranteed 5 minute 
delivery time.  The e-mail delivery is orders of magnitude faster than surface 
mail, but that does not necessarily mean it qualifies as a real-time delivery 
system.  It is the required delivery deadline that ultimately establishes 
whether the real-time system definition has been met.  For example, both systems are 
real-time systems if the deadline is six days because both are logically 
and temporally correct.  However, neither one is a real-time system if the 
deadline is three minutes because neither one is temporally correct. 

Real-time systems are classified as hard or soft.  Hard systems have 
catastrophic consequences if the temporal requirements are not 
met---up to and including complete system destruction.  In fact, if the hard system is 
safety-critical, failure could lead to injury or even death. Conversely, soft 
systems only have degraded performance if the temporal requirements 
aren't met.  The classification of a fault-tolerant system, in particular, 
depends on the nature of the faults and the consequences for failing to 
detect and correct them in a timely manner.  Suppose a fault results in an 
over-temperature condition.  If the system hardware can survive this 
condition for up to five minutes, then fault recovery must be completed 
within five minutes to prevent further damage.  This would be a hard 
fault-tolerant system.  On the other hand, if the fault only causes a 
minor loss of some sensor data, fault recovery could take considerably 
longer without dire consequences.   This would be a soft fault-tolerant 
system.  

\section{Quantifying Reconfiguration Time} \label{s3}

\begin{table*}[ht]
\begin{tabular}{|c|c|c|c|c|c|c|} \hline
\bf{Device} & \bf{Type} & \bf{Size} & \bf{Mfg} & $\mathbf{\tp}$ (ms) & \bf{Ref.} & \bf{Notes}\\ \hline \hline
ispPAC10 & FPAA & 4 & Lattice Semiconductor & 100  & \cite{pac10} & \\ \hline
AN220E04 & FPAA & 4 & Anadigm  & 3.8 & \cite{fpaa0} & 1, 2\\ \hline
JPL's FPTA2 & FPTA & 64 & fabricated by MOSIS & 0.008 & \cite{jpl22} & 3 \\ \hline
\end{tabular} 
\vspace{1ex} 

(1) All 18 banks are reloaded with 256 bytes/bank \\
(2) Serial transfer with 10 MHz clock \\ 
(3) Byte-wide transfers with 160 MHz clock  
\caption{Programming times for various popular reconfigurable analog devices.  
All are COTS devices except for the FPTA.  The units for size are 
modules for FPAAs and cells for FPTAs. The references indicate 
where the $\tp$ value is documented.} 
\label{tbl:1}  
\end{table*}

\vspace{1ex}

The main parameter we concentrate on is the programming time ($\tp$) for 
reconfigurable analog devices.  
Table \ref{tbl:1} shows the programming time ($\tp$) for several 
reconfigurable devices. 
This programming time cannot be ignored
because EHW algorithms frequently have 
populations sizes in the hundreds and they run for thousands of 
generations.

\vspace{1ex}

\noindent\textit{Example 1:}

\vspace{1ex}

Suppose a proportional-derivative 
controller is implemented in an FPAA.  
A controller's fitness is found by applying a step
input to the control system and then measuring its 
settling time.  The fitness evaluation lasts at least as
long as the settling time does, which can be somewhat lengthy.
Indeed, settling times of two minutes are not unheard of~\cite{ngst}.  
Under these circumstances, it wouldn't 
take a very large population size nor a large number of generations 
to make an intrinsic reconfiguration run for hours or even
days before finishing.

\vspace{1ex}

\noindent\textit{Example 2:}

\vspace{1ex}

An AN220E04 FPAA is used to compensate for aging effects in a 
control system responsible for positioning a satellite's communications 
antenna. The reconfiguration search is done by a generational GA run 
for 500 generations with a population size of 100.  The system's step 
response is measured to determine if the compensation is correct.  This 
step response test takes 625 milliseconds to conduct.  It takes 
3.8 + 625 = 628.8 ms to reprogram the FPAA and compute the evolved compensator's 
fitness, but a total of 500,000 compensators are evolved during the evolutionary
algorithm's run.  Hence, the reconfiguration time takes $\approx 8.7$ hours.

\section{Discussion} \label{s5}

Reconfiguration times are meaningless until they are put into context.
For instance, take Example 2 from the previous section.  
Suppose brief communication sessions with the satellite 
are scheduled at 10 hour intervals.  A session may be skipped, but skipping two 
sessions in a row is not permitted.  If a fault is detected just prior to a 
scheduled session, and if the error results in missing the session, then the 
fault recovery deadline is 10 hours.  This is the worst case scenario\footnotemark.  
An almost 9 hour reconfiguration time may seem quite long, but in this case it is 
perfectly acceptable because $T_r < 10$.  On the other hand, it would 
not be acceptable if communication sessions were scheduled at 6 hour intervals.

\footnotetext{~Missing one session is permitted.  If the fault is detected 
just after a scheduled session, the fault recovery deadline
would be 20 hours.}

The only way to determine if there is a problem is to compare the reconfiguration 
time against the fault recovery deadline.  This latter quantity is system dependent.  
No problem exists so long as the reconfiguration time is less than the recovery deadline.

This time comparison adds a new perspective on intrinsic evolution and, at the same 
time, imposes a new requirement.  Reconfiguration becomes a real-time process whenever
it is used as a fault recovery method.  Consequently, it is no longer sufficient 
to just talk about how an evolutionary algorithm was able to restore a circuit's 
functionality.  These statements may show logical correctness, but without comparing the
reconfiguration time against a deadline there is no proof of temporal correctness.  
Just reporting an algorithm's running time doesn't say anything about temporal correctness either.
The key point is expressed by the following first principle:  

\begin{quote}
\emph{
No EHW-based recovery method can legitimately proclaim 
efficacy until it is proven to be both logically and temporally correct.}
\end{quote}

The validity of this principle is easy to see.  If the recovery 
method isn't logically correct, then the problem can't be fixed.  If it 
isn't temporally correct, then the problem can't be fixed soon enough
to prevent other things from going wrong.  
Without proving logical \underline{and} 
temporal correctness, there is no basis
for claiming a fault recovery method is effective.

It is easy to prove if a fault recovery method is logically 
correct---try it and see if it fixes the problem.  Proving temporal 
correctness, however, is  more complicated because it really depends 
on conducting a thorough failure modes and effects analysis
(FMEA).  This analysis should identify all potential faults and their 
effects on system performance~\cite{fmea}.  One outcome of a FMEA are 
the recovery deadlines.  Temporal correctness is proven if a logically 
correct recovery is guaranteed to finish prior to the recovery 
deadline.  

Greenwood, et.~al~\cite{gree} suggested evolutionary algorithms 
designed for reconfiguration searches perform best if they have high
selection pressure and if they emphasize mutation for reproduction.
In principle, any type of evolutionary algorithm could be used for 
a reconfiguration search, but from a practical standpoint genetic 
programming algorithms should be avoided.  Genetic programming 
algorithms designed for EHW problems are put on large multiprocessor 
systems to abridge their long running time~\cite{gpsks2,gpsks1}.  
This becomes especially problematic for fault-tolerant systems because, 
if there isn't enough room for redundant hardware, then there isn't 
enough room for a large multiprocessor system either.  It seems 
unlikely a full-fledged genetic programming search,
run on a single processor, could finish quickly enough to meet 
a fault recovery deadline of only a few hours.

\section{Conclusions}

EHW-based reconfiguration is a viable method of performing fault 
recovery in systems without redundant hardware.  Fault-tolerant 
systems are real-time systems.  Consequently, any attempts to 
intrinsically evolve a new hardware configuration must consider 
the device programming time and the fitness evaluation time 
because they both contribute
to the reconfiguration time.  

It has been shown neither a large population size nor thousands 
of generations are necessary to have reconfiguration searches 
with surprisingly long finishing times.  
However, a long search time by itself  is 
not enough to reject reconfiguration as a fault recovery method. 
Intrinsic reconfiguration can be used for fault recovery so long as
it finishes before the mandatory recovery deadline.

\bibliography{fault}

\end{document}